# Dual-wavelength Photo-Hall effect spectroscopy of deep levels in high resistive CdZnTe with negative differential photoconductivity

A. Musiienko[1,a)], R. Grill[1], P. Moravec[1], G. Korcsmáros[1], M. Rejhon[1], J. Pekárek[1], H. Elhadidy[2,3], L. Šedivý[1] and I. Vasylchenko[1]

[1]*Charles University, Faculty of Mathematics and Physics, Institute of Physics, Ke Karlovu 5, CZ-121 16, Prague 2, Czech Republic*

[2]*Central European Institute of Technology, Institute of Physics of Materials, ASCR, Brno 61662, Czech Republic*

[3]*Faculty of Science, Physics Department, Mansoura University, Mansoura 35516, Egypt*

## ABSTRACT

Photo-Hall effect spectroscopy was used in the study of deep levels in high resistive CdZnTe. The monochromator excitation in the photon energy range 0.65–1.77 eV was complemented by a laser diode high-intensity excitation at selected photon energies. A single sample characterized by multiple unusual features like negative differential photoconductivity and anomalous depression of electron mobility was chosen for the detailed study involving measurements at both the steady and dynamic regimes. We revealed that the Hall mobility and photoconductivity can be both enhanced and suppressed by an additional illumination at certain photon energies. The anomalous mobility decrease was explained by an excitation of the inhomogeneously distributed deep level at the energy $E_v+1.0$ eV enhancing thus potential non-uniformities. The appearance of negative differential photoconductivity was interpreted by an intensified electron occupancy of that level by a direct valence band-to-level excitation. Modified Shockley-Read-Hall theory was used for fitting experimental results by a model comprising five deep levels. Properties of the deep levels and their impact on the device performance were deduced.

---

[a] Electronic mail: musienko.art@gmail.com

# 1. Introduction

An engagement of semiconductors in electronic applications is considerably controlled by deep levels (DLs) associated with crystallographic imperfections[1,2]. Such defects are responsible for the recombination processes[3], space charge formation and polarization of biased sample[4,5], and emergence of potential non-uniformities inside the device[1,3]. CdTe and CdZnTe crystals have a wide variety of applications in X-ray and gamma-ray detection[6,7], photovoltaics[8], security[9], medicine[10], and investigation of the universe[11]. The material quality strongly controls the price of the ingots[7] and further investigation and improvement of the material are therefore needed. The existing thermal emission spectroscopy methods suffer from known limitations when unambiguously identifying DLs properties, especially in case of very deep levels with trapping energy well above the half of the band gap energy[12,13]. In our previous paper[13] we showed that photo-Hall effect spectroscopy (PHES) is a convenient method for the complementary study of deep level properties. The Hall mobility $\mu_H = \sigma R_H$ determined by PHES experiment, where σ, $R_H$ are the conductivity and Hall coefficient, respectively, is an excellent parameter that may be used for identification of structural imperfections and inhomogeneity inducing electric potential non-uniformities in the material[14]. The appearance of such disorder leads to an apparent variations of $\mu_H$[15,16] and may be easily identified by PHES.

The aim of this paper is to extend previous effort focused on investigating properties of deep levels in semi-insulating CdZnTe and reveal the connection between the appearance of particular features in measured quantities and the structure of deep levels in the sample. We report on a more elaborated PHES with the simultaneous dual-wavelength illumination (DWPHES) where an additional extensive sub-bandgap laser diode (LD) illumination allows us to reveal the important influence of the DLs and enhance PHES signal. We present a complex analysis of DWPHES measurements and develop deep levels model supported by consistent theoretical calculations. We argue that this approach offers a more reliable determination of deep level parameters than other recently developed methods[17–20].

## 2. Methods

## 2.1 Sample properties and photo-Hall effect measurements

For this contribution, a single crystalline CdZnTe with a zinc content of 4.5 % grown by the vertical gradient freeze method at the Institute of Physics, Charles University was chosen. N-type sample with dimensions of 3x2x12 mm$^3$ prepared by a standard method[13] was used in the classic six-contact Hall-bar shape convenient for galvanomagnetic measurements. The sample was distinguished by an unusual decrease of mobility, the appearance of negative differential photoconductivity (NDPC) and several DLs. The Fermi energy $E_F = E_C - 0.64$ eV was fixed in accord with given n-type resistivity of 9x10$^8$ Ωcm. The band gap energy $E_g = 1.55$ eV was used in calculations. Another sample from this crystal adapted as radiation detector revealed fast polarization[21]. Other data on this sample were presented in Ref[13] for sample labeled No. 1.

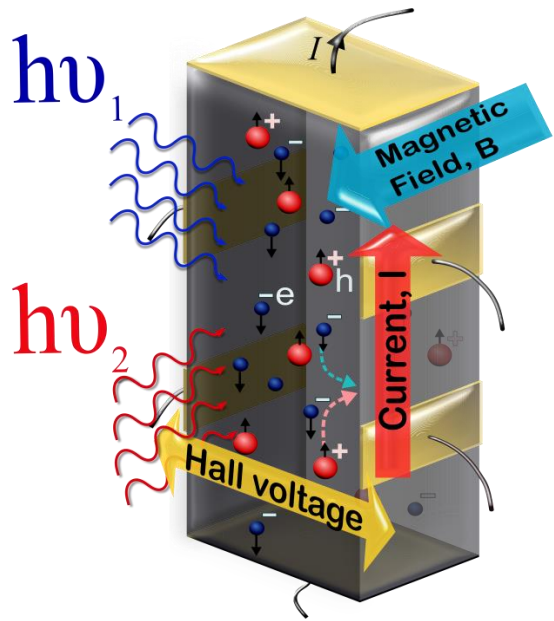

**Fig. 1.** Basic principles of DWPHES measurements.

Basic principles of photo-Hall effect spectroscopy with dual-wavelength illumination are shown in Fig. 1. The Hall voltage produced by 1T magnetic field is measured under simultaneous illumination from the two light sources. One of the sources, a 100 W halogen lamp filtered through a monochromator with the maximum photon flux of $7\times10^{13}$ cm$^{-2}$s$^{-1}$, has tunable photon energy. The second source has fixed photon energy. For this purpose, we chose 0.8, 0.95, or 1.27 eV Thorlabs Laser Diodes with the photon flux up to $1.4\times10^{16}$ cm$^{-2}$s$^{-1}$. Fiber wavelength mixer is used to combine the two illumination sources and to provide homogeneous illumination of the sample. All measurements were performed at room temperature.

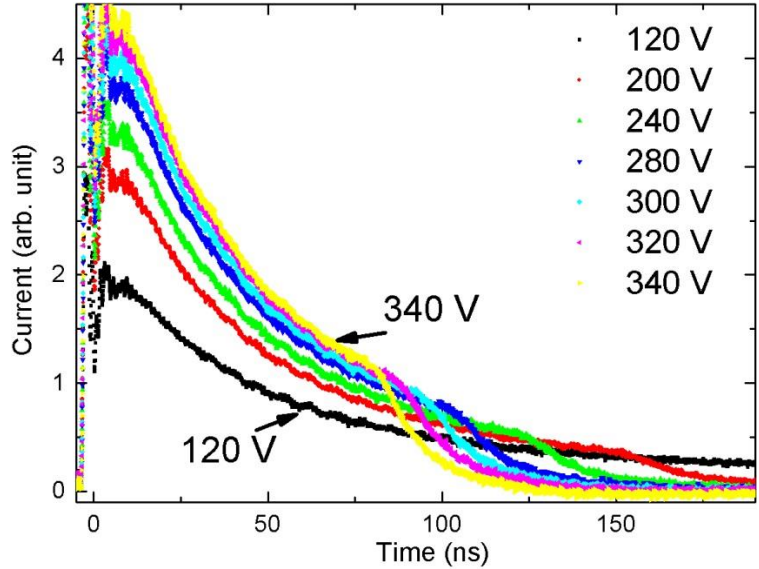

**Fig. 2.** Laser-induced transient current technique waveforms measured with 5 ms voltage pulse at different biasing.

The deep level threshold energies were identified from DWPHES by monitoring $\mu_H$ and photoconductivity (PhC) spectra. The electron lifetime, $\tau_e \approx 120$ ns was measured by a laser induced transient current technique (L-TCT[22]) in a pulsed bias regime where 5 ms voltage pulse allowed us to overcome the polarization of the device, see Fig. 2. Rather low electron mobility-lifetime product $\mu_e \tau_e \approx 10^{-4}$ cm$^2$V$^{-1}$ was determined by the Hecht equation fit according to Uxa et al.[23].

## 2.2 Charge dynamics model

The Shockley-Read-Hall (SRH) theory[24] completed by illumination-mediated deep level - band transitions is used for the evaluation of charge dynamics. Equations (1-5) represent correspondingly the electron, hole and trapped electron dynamics. Electron and hole recombination-generation rates are described by eq. 4 and 5:

$$\frac{\partial n}{\partial t} = -\sum_i U_i^e, \qquad (1)$$

$$\frac{\partial p}{\partial t} = -\sum_i U_i^h \qquad (2)$$

$$\frac{\partial n_{ti}}{\partial t} = U_i^e - U_i^h. \qquad (3)$$

$$U_i^e = I\tilde{\alpha}_{ei} n_{ti} - \alpha_{ei} v_e [(N_{ti} - n_{ti})n - n_{ti} n_{1i}] \qquad (4)$$

$$U_i^h = I\tilde{\alpha}_{hi}(N_{ti} - n_{ti}) - \alpha_{hi} v_h [n_{ti} p - (N_{ti} - n_{ti}) p_{1i}] \qquad (5)$$

Here $n$, $p$, $n_{ti}$, and $U_i^{e(h)}$ are the densities of free electrons, free holes, electron trapped in the $i$-th level, and electron (hole) net recombination rate at the $i$-th level, respectively. The quantities defining recombination rates $n_i$, $N_{ti}$, $\alpha_{ei}$, $\alpha_{hi}$, $v_e$ and $v_h$ in eq. (4-5) are intrinsic carrier density, $i$-th deep level density, electron and hole thermal capture cross-sections, and electron (hole) thermal velocities, respectively. Symbols $n_{1i}$ and $p_{1i}$ stand for electron and hole densities in case of $E_F$ being set equal to the deep level ionization energy $E_{Ti}$[24]. The effect of the illumination on the $i$-th DL occupancy is defined by the photon flux $I$ and photon capture cross-sections relevant to the conduction and valence band transition by $\tilde{\alpha}_{ei}$, and $\tilde{\alpha}_{hi}$. The order of magnitude of $\tilde{\alpha}_{e(h)i}$ was estimated from Ref.[18] and rises appropriately with the PhC spectra values. The charge neutrality is assured by the neutrality equation

$$p - n - \sum_i n_{ti} = p_0 - n_0 - \sum_i n_{t0i}, \quad (6)$$

where the equilibrium zero-indexed quantities on the right-hand side are defined by the position of $E_F$. The solution of equations (1)-(3) significantly simplifies in the steady state regime, where the time derivatives are set to zero.

## 3. Results and Discussion

### 3.1 Experimental results

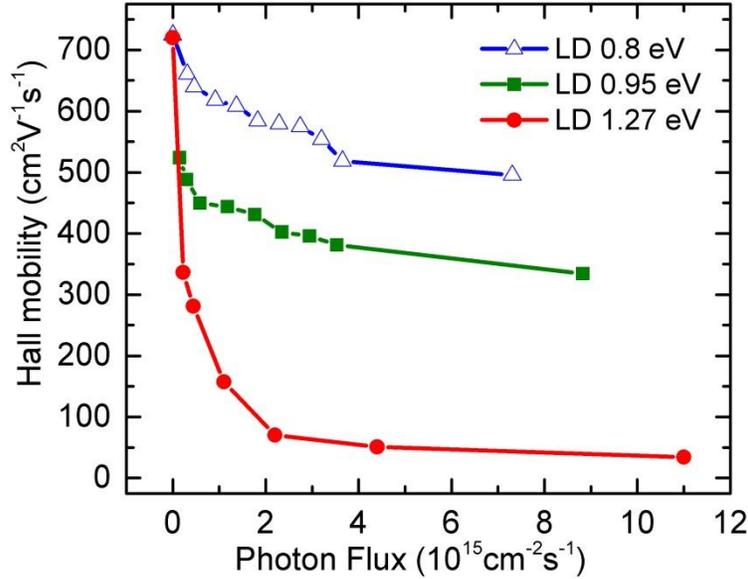

**Fig. 3.** Hall mobility versus laser induced photon flux at 0.8, 0.95 and 1.27 eV photon energy.

In contrast to common high resistive detector grade CdZnTe and CdTe samples, where the illumination habitually induces an enhancement of the Hall mobility[25], chosen CdZnTe sample subjected to an extensive illumination showed an anomalous decrease of $\mu_H$, see Fig. 3. Further investigation of the sample showed that PhC intensity dependencies (PhC $\propto I^\alpha$) reveal both superlinear regions[26] with $\alpha > 1$ at low photon flux $< 10^{13}$ cm$^{-2}$s$^{-1}$ at 1.0 eV photon energy and sublinear regions[27] with $\alpha < 1$ at energies of 0.7, 0.8, 0.95, and 1.27 eV, see Fig. 4. Negative differential photoconductivity[28] was observed at 1.27 eV LD illumination with the photon flux above $5\times10^{14}$ cm$^{-2}$s$^{-1}$. Simultaneously, an increasing noise of the signal represented by error bars in Fig. 4 (a) appeared.

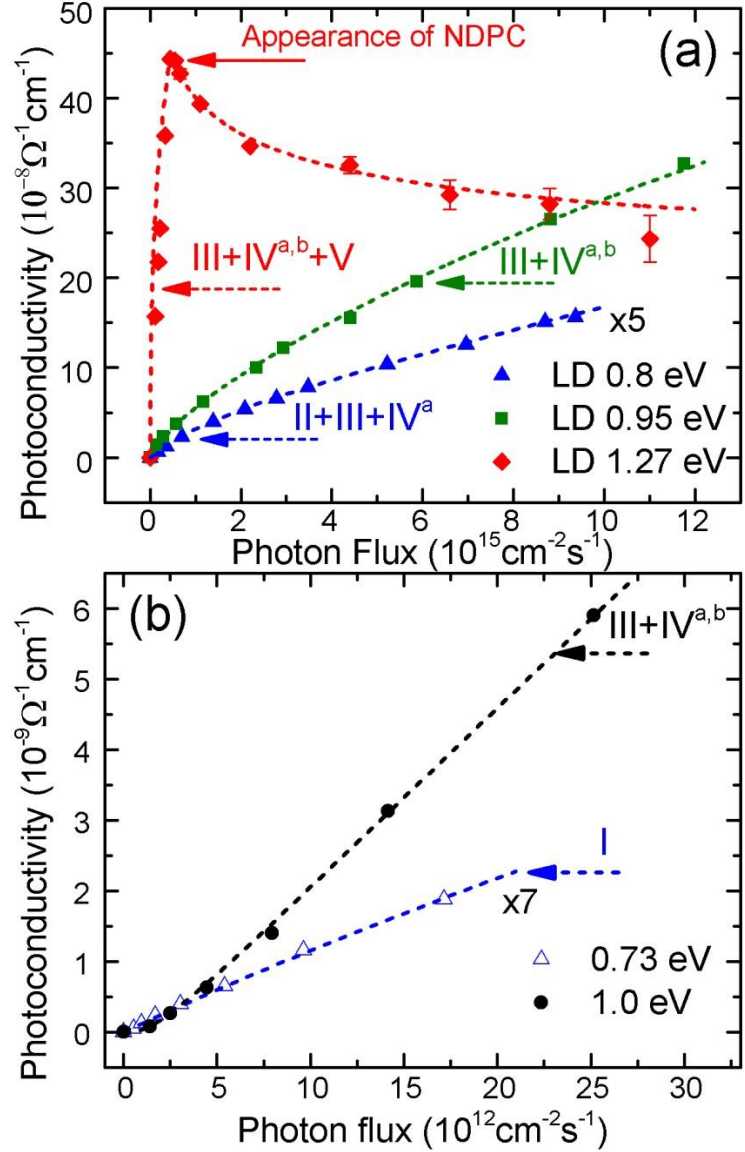

**Fig. 4.** PhC as a function of the extended (a) and low (b) photon flux at 0.73, 0.8, 0.95, 1.0 and 1.27 eV photon energies (hυ). The dashed curves show the fit with the SRH model discussed in the text with parameters in Table 1. The error bars appear only if standard deviation exceeds a symbol size. Roman numerals designate respective transitions defined in the deep levels model responsible for photoconductivity rise produced by free carriers generation.

To discover the reasons of the above-mentioned effects as well as low $\mu_e\tau_e$ and fast detector polarization[21], we have performed extensive DWPHES measurements in various regimes of simultaneous single and two-photon illumination. Fig. 5 shows $\mu_H$ without and with an additional laser diode illumination. One can see threshold points near 0.75, 0.9, 1.0 and 1.2 eV in the spectra for all illumination regimes. In case of single wavelength illumination shown in Fig. 5(a), $\mu_H$ increases at 0.75 eV, 0.9 eV, and 1.27 eV photon energies. A significant decrease of $\mu_H$ can be observed after $h\upsilon > 1.0$ eV,

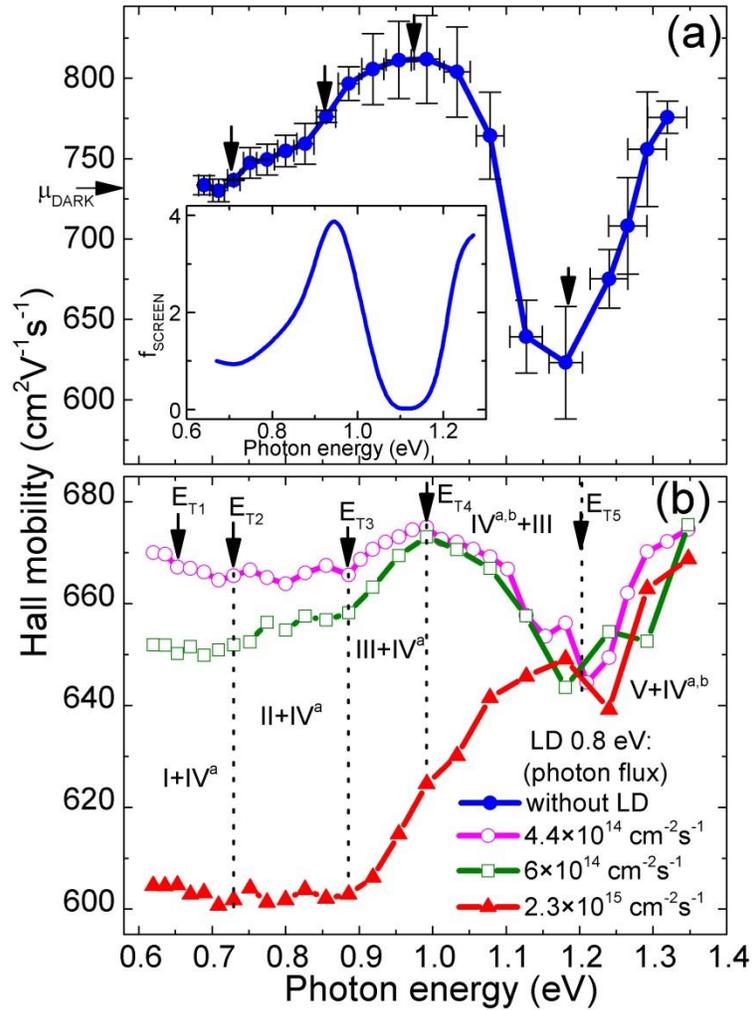

**Fig. 5.** Hall mobility spectra. Upper panel (a) shows PHES obtained without laser diode illumination. Bottom panel (b) gives plots of $\mu_H$ obtained by simultaneous illumination with monochromator and 0.8 eV laser light of different intensity. Vertical arrows indicate deep level threshold energies. The vertical

and horizontal error bars represent standard deviation and monochromator experimental error, respectively. Roman numerals indicate energy regions representing different generation-recombination processes discussed later on. The inset picture shows a screening coefficient.

region IV[a,b]+III. Similar effects are also seen after additional laser diode 0.8 eV illuminations, where in addition $\mu_H$ is suppressed below 0.9 eV. The spectrum without laser diode illumination appeared noisier than that with laser diode, as denoted by vertical error bars in Fig. 5(a). Stabilization of the Hall voltage by additional laser illumination can be the keystone of this method, which can allow the Hall signal measurements not only in detector grade samples but also in the poor quality ones.

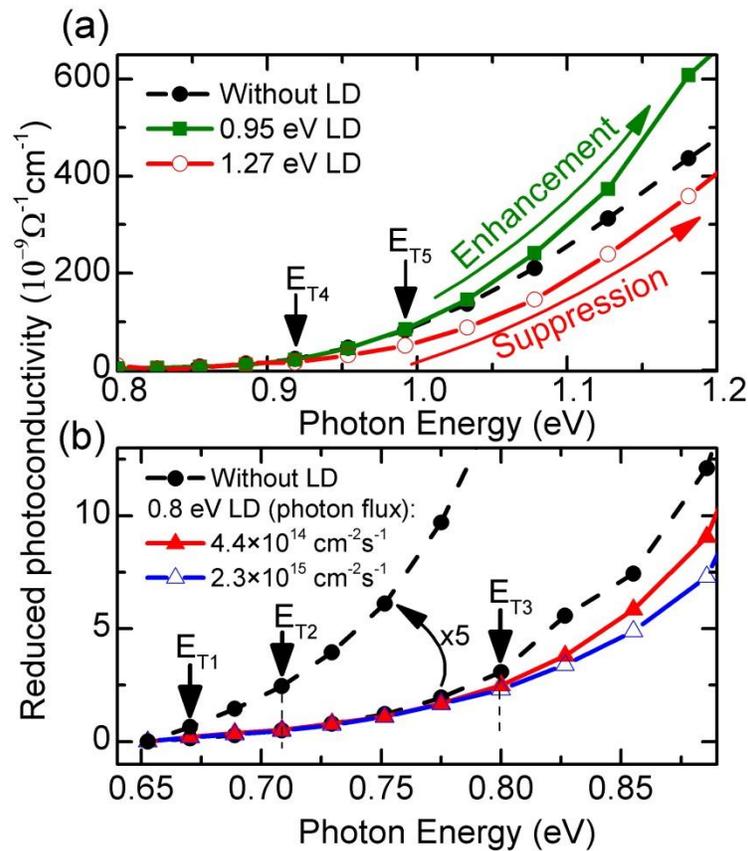

**Fig. 6.** Reduced photoconductivity spectra with the additional LD illumination at (a) 0.95 eV, 1.27 eV and (b) 0.8 eV. Black dashed curves with full bullets show spectra without additional illumination.

Corresponding LD intensities were chosen to show best the effect of reduced PhC enhancement and suppression.

A comparison of conventional PhC, $\sigma_M$, and PhC with additional laser diode illumination may be used for the identification of DLs position relative to respective bands. With this aim, we define a reduced photoconductivity $\sigma_R = \sigma_{M+LD} - \sigma_{LD}$, where $\sigma_{M+LD}$, and $\sigma_{LD}$ are the photoconductivities under simultaneous monochromator and laser illumination, and PhC produced by purely laser illumination, respectively. In this way $\sigma_R$ and $\sigma_M$ can be compared on the same scale. Respective results are summarized in Fig. 6. While a suppression of $\sigma_R$ is observed in case of PhC after additional 0.8 eV or 1.27 eV laser diode illumination as an obvious consequence of the depletion of related DLs labeled $E_{T2}$, $E_{T3}$, and $E_{T5}$ in the defect model, the 0.95 eV laser diode illumination gives an enhancement of $\sigma_R$. We interpret this fact by a synergism of the excitation processes involving $E_V$ to $E_{T4}$ level transition at the 1.0 eV energy and the $E_{T4}$ to $E_C$ transition. Consequently, while DLs energies $E_{T2}$, $E_{T3}$, and $E_{T5}$ are counted relative to transitions). The photon energies represent threshold energies defined by experiment. In case of upward arrows, Roman numerals show electron (hole) generation process activated by the photon energy $hv>E_T$. A downward arrow with IV$^a$ numeral represents electron capture process on the $E_{T4}$ level associated with potential non-uniformities. Same Roman numerals are used in the discussion to indicate processes involved in generation (recombination) of free carriers.

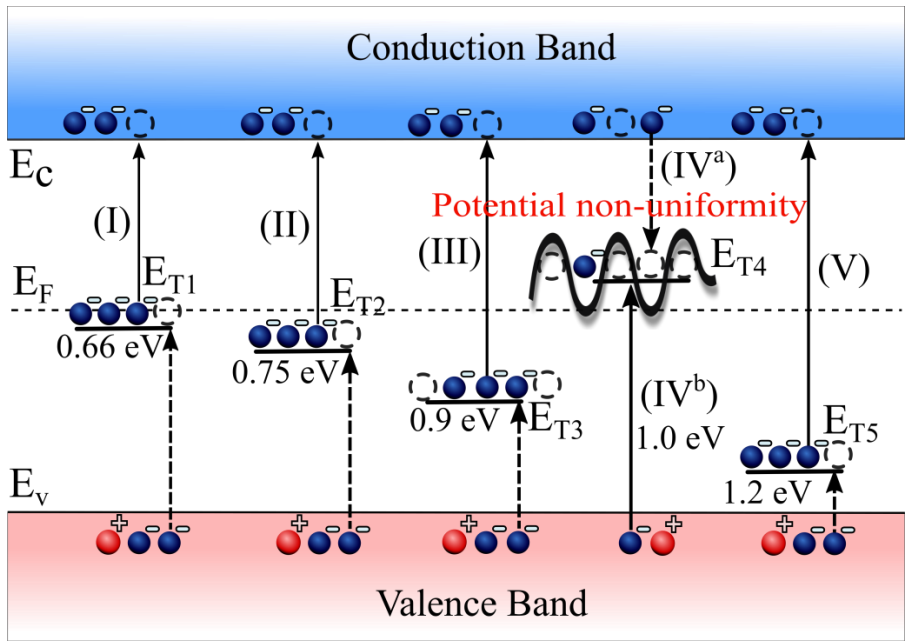

**Fig. 7.** The defect model with pondered transitions of the DLs. Full upward arrows delineate principal optical excitation, dashed arrows show thermally activated transitions (or supplementary optical

the conduction band, $E_{T4}$ is expressed relative to the valence band. It is important to note here that another possibility of deep level positions in the band gap fails. The superlinear behavior, the same as the enhancement of the reduced photoconductivity at 0.95 eV, can't be reached by the level below Fermi energy. Such level must create positive charged potential non-uniformities but in the n-type material the concentration of the minority holes is negligible. Hence this possibility is unlikely. Less impact from these DLs transitions is also observed in Fig. 5, where the value of $\mu_H$ in the region II+IV$^a$ is suppressed and maximum rises of $\mu_H$ in the regions III+IV$^a$ and V+IV$^{a,}$b are smaller in comparison with the spectra without laser diode illumination. The impact of DL $E_{T1}$ on PhC remains untouched, mainly because of the nearly linear character of PhC vs photon flux apparent in Fig. 4(b) at 0.73 eV. The position of this level may be uniquely allocated relative to $E_C$ as it is observed below the middle of the band gap excitation, where an opposite transition could not influence the electron density[13].

## 3.2 Shockley-Read-Hall model simulation results

Experimental results presented in Figs. 2-6 completed by given $\mu_e\tau_e$ were incorporated into the simulations within the five-level scheme outlined in Fig. 7. Parameters $\alpha_{e2} = 2\times10^{-17}$ cm$^2$, $\alpha_{e3} = 1.7\times10^{-17}$ cm$^2$, $\alpha_{e5} < 10^{-16}$ cm$^2$ were estimated from PhC transients measured on this sample[13]. For simplicity, deep level $E_{T4}$ was considered as the level of an effective density $N_{T4}$. The fit of PhC is shown by dashed lines in Fig 4 where PhC was defined as PhC = $e\mu_H(n-n_0)$. We treated the Hall mobility, concentration, and photoconductivity as effectively representing mean values through the sample. The electron mobility in Fig. 3 was used in the fit. First we found superlinear and sublinear behaviors of PhC that give limitation on parameters of the deep levels $E_{T3}$ and $E_{T4}$, Fig 4(b) curves LD 0.95 and 1.0 eV. Secondly, the shape of NDPC shoulder was obtained at extended 1.27 eV illumination, which gives the parameters of the level $E_{T5}$. The $\alpha_e \cdot N_{T1}$ product can be estimated from the electron lifetime obtained from pulsed L-TCT. The last step was the variation of the parameters to obtain a qualitative fit with all of the curves simultaneously. The parameters resulting from the fit are given in Table 1. Due to the lack of information about the levels $E_{T1}$ and $E_{T2}$ some parameters can vary over a wide range without an influence on the fit. Those values are not presented in the Table.

**Table 1.** Parameters of the DLs model.

| DL | $E_{T1}$ | $E_{T2}$ | $E_{T3}$ | $E_{T4}$ | $E_{T5}$ |
|---|---|---|---|---|---|
| Position in the bandgap, eV | $E_c - 0.66$ | $E_c - 0.75$ | $E_c - 0.9$ | $E_V + 1.0$ | $E_c - 1.2$ |
| $N_t$, cm$^{-3}$ | $10^{14}$ | $10^{12}$ | $10^{14}$ | $3 \times 10^{13}$ | $10^{15}$ |
| $\alpha_e$, cm$^2$ | $3 \times 10^{-15}$ | $2 \times 10^{-17}$ | $1.7 \times 10^{-17}$ | $4 \times 10^{-20}$ | $10^{-14}$ |
| $\alpha_h$, cm$^2$ | $10^{-15}$ | -* | $10^{-17}$ | $10^{-18}$ | $8 \times 10^{-15}$ |
| $\tilde{\alpha}_{ei}$, $10^{-17}$cm$^2$ | 470** | 5 | 2.9 | 16 | 30 |
| $\tilde{\alpha}_{hi}$, $10^{-17}$cm$^2$ | - | 7 | 22 | 2.4 | 10 |
| PhC character | Sub.*** | Sub. | Sub. | Sup. | Sub. |

*The values can vary over a wide range without an influence on the fit.

**The values of $\tilde{\alpha}_{ei}$ and $\tilde{\alpha}_{hi}$ are given correspondingly at the 0.73, 0.8, 0.95, 1.0, and 1.27 eV illumination energy.

***Sub./Sup.=Sublinear/Superlinear.

SRH model simulations of the photoconductivity spectra with single and dual illumination regimes with parameters from the Table 1 are presented in Fig. 8. One can see that the presence of the deep level in the energy region results in a change of the envelope of the curve and rise of PhC. The suppressed enhanced curves obtained by additional illumination follow a similar course as the experimental spectra in Fig. 6. Despite a good correlation, theoretical curves have a slightly different shape in the region $hv > 1.0$ eV. This can be explained by increased absorption in the region near the Urbach tail which is commonly observed in CdZnTe[29].

A prominent achievement was reached in the identification of deep level responsible for the drop of $\mu_H$ presented in Fig. 5. This effect is explicitly joined with the excitation of the deep level $E_V+1.0$ eV labeled by IV in Fig. 7. Deep level $E_{T4}$ is assumed to be acceptor so that it charges negatively with the electron filling.

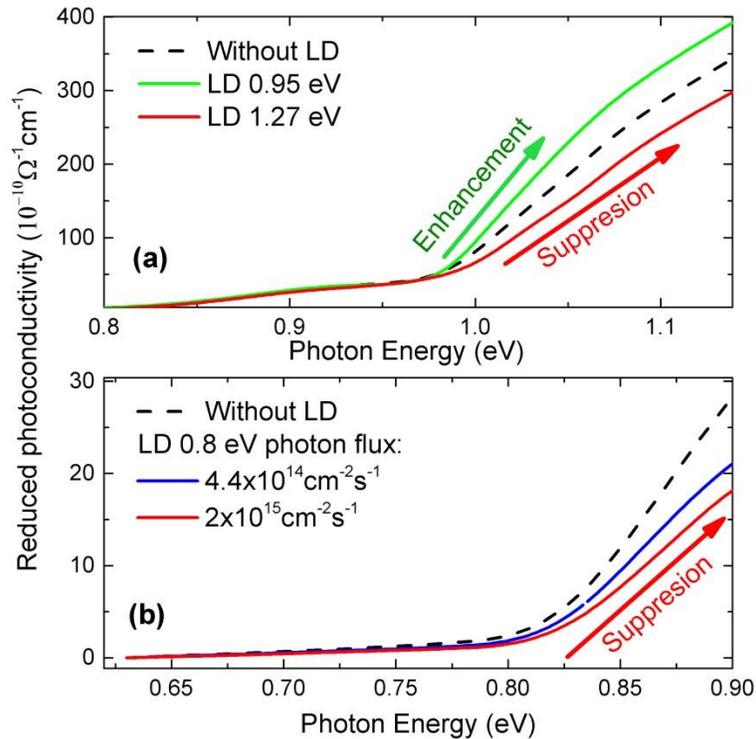

**Fig. 8.** Reduced photoconductivity spectra with the additional laser diode illumination at (a) 0.95 eV, 1.27 eV and (b) 0.8 eV obtained by SRH theoretical simulation with parameters from Table 1. Black dashed curves show theoretical spectra without additional illumination.

Electrons occupying the deep level $E_{T4}$ produce negatively ionized charge (Fig. 7 IV$^a$ and IV$^b$) and create Coulomb potential non-uniformities[30] (CPNUs). We explain such correlation by a nonhomogeneous spatial distribution of defects on this level. Its charging entails a formation of CPNUs[31], which consequently influence the carriers transport[32]. We assume that the bulk of the sample may be divided into two regions as outlined in Fig. 9 . Region 1 contains the level $E_{T4}$ with abundant density, in contrast to region 2 with $E_{T4}$ level suppressed density. While the level $E_{T4}$ is nearly empty in the dark and related CPNUs and bands warping are small, the optical excitation results in the level $E_{T4}$ filling and due to its nonhomogeneous distribution an CPNUs increase entailing observed $\mu_H$ reduction. Relevant model of the Hall and photo-hall effects in inhomogeneous materials was theoretically elaborated in Ref[15] and our findings agree well with that concept, see Fig. 3 in Ref[15]. Let us note that the explanation of the $\mu_H$ depression by other contingent models fails. The $\mu_H$ limitation caused by the enhanced ionized impurity scattering is irrelevant at room temperature and ionized defect density significantly below $10^{17}$ cm$^{-3}$, where $\mu_H$ is dominantly limited by the optical phonon scattering[33]. Similarly, the effect of holes on the $\mu_H$ reduction may be excluded due to the character of DWPHES signal, where an indication of sign conversion was never detected.

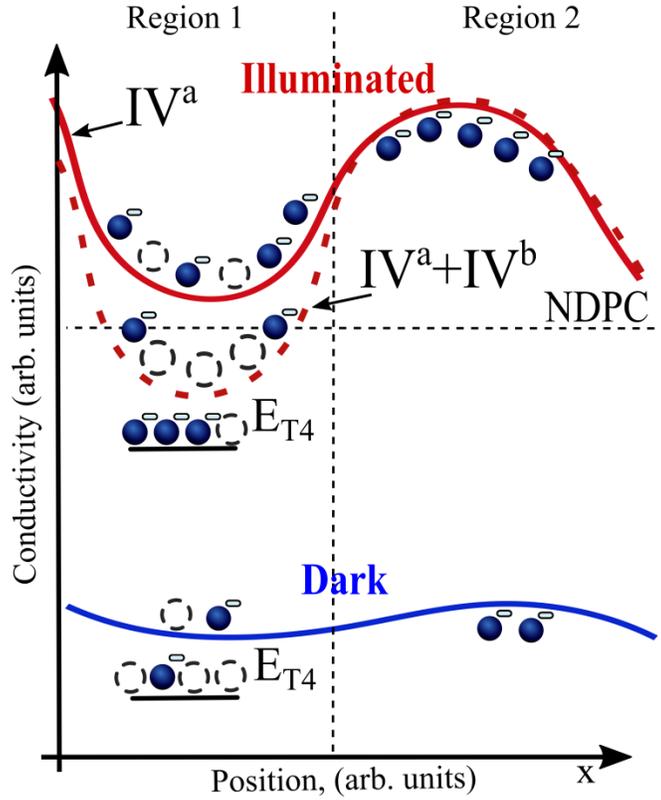

**Fig. 9 .** Schematic diagram of nonhomogeneous conductivity distribution in the sample at different illumination regimes. Region 1 and Region 2 present parts of the sample with and without deep level $E_{T4}$. Electrons and empty circles show a relative change of carrier concentration in the conduction band and on the DL $E_{T4}$. Roman numerals indicate processes from Fig. 7 which lead to the occupation of the level $E_{T4}$. Dashed line outlines the scheme of NDPH induced by relative electron depletion in Region 1 at high excitation intensity.

The enhancement of the Hall mobility occurs as a result of the compensation or screening of CPNU[30] by producing compensating charge on other DLs: $E_{T1}$, $E_{T2}$, $E_{T3}$ and $E_{T5}$. The mechanism may be visualized by defining a screening coefficient $f_{SCREEN}=n^*/n_{t4}$, which weighs the screening of the Coulomb interaction expressed by an effective concentration of the screening charge[34] $n^* = n + p + \sum_1^5 n_{ti}(1 - n_{ti}/N_{ti})$ and by the charge density $n_{t4}$ on the $E_{t4}$ DL responsible for CPNUs. Increased $n^*$ enhances $\mu_H$, which is damped by the population of $E_{t4}$ DL. The shape of $f_{SCREEN}$ follows well the course of $\mu_H$, see the inset in Fig. 5(a) The mobility enhancement in regions II+IV$^a$, III+IV$^a$ and V+IV$^{a,b}$ points to the presence of the DLs with threshold energies $E_C - E_{T2} = 0.75$ eV, $E_C - E_{T3} = 0.9$ eV, and $E_C - E_{T5} = 1.2$ eV. The $\mu_H$ depression at the high intensity LD 0.8 eV illumination apparent in Fig. 5(b) is explained by the $E_{T4}$ population via a transfer of electrons from $E_{T1}$ or $E_{T2}$ DLs according to the scheme $E_{T(1)2} \Rightarrow E_C \Rightarrow E_{T4}$ and by absorption of two photons by an excitation from the valence band through $E_{T1}$ or $E_{T2}$ DLs $E_V \Rightarrow E_{T1(2)} \Rightarrow E_C \Rightarrow E_{T4}$. Let us note that weak depletion of DLs localized below $E_F$ results in an enlarged screening while a significant depletion of these levels with $n_{ti} < N_{ti}/2$ damps the screening oppositely.

### 3.3 Negative differential photoconductivity

Negative differential photoconductivity is characterized by the decrease of the photoconductivity under illumination[35]. The effect is mathematically described by a negative derivative of photoconductivity *PhC* with respect to illumination intensity *I*. It can be induced both by a decrease of the carrier concentration or by a decrease of the carrier mobility according to the formula

$$\frac{dPhC}{dI} = e(\mu \frac{dn}{dI} + n \frac{d\mu}{dI}), \tag{8}$$

where only electrons as majority carriers were involved. As it was shown in Ref[15], the distinction of variations of *n* or *μ* is distinguishable in inhomogeneous materials and an attempt to derive right *n* or *μ* spatial distribution based on Hall data is not accessible. The appearance of NDPC points to abnormality in the sample's defect structure leading to redundant carrier recombination and deterioration of the detection properties of CdZnTe. The negative differential photoconductivity has not yet been reported in CdZnTe. In our measurement the NDPC was not observed at the energies $hv < E_{T4}$, see Fig. 4. At the same time the Hall mobility decreases monotonically at all extended fluxes, see Fig. 3. This effect can be explained by two different processes, which lead to filling of the level $E_{T4}$. The process IV[a], see Fig. 7, dominates at the $hv < E_{T4}$ and the increase of the free carrier concentration has a stronger influence on the photoconductivity than the decrease of mobility. The second generation channel is activated at $hv > E_{T4}$ photon energies, which results in the increased DL $E_{T4}$ filling. Meanwhile, deep levels below

Fermi energy are largely depleted from electrons due to intensive photo-excitation and the electron density decreases in Region 1, see Fig. 9. These two simultaneous events are described schematically in Fig. 9 by the transition labels (IV$^a$ + IV$^b$) defined in Fig.7. Interestingly, the photon capture probability of process IV$^a$ is higher than of IV$^b$ channel, $\tilde{\alpha}_{e4} > \tilde{\alpha}_{h4}$ see Table 1. This results in the increase of PhC at relatively low photon fluxes ($I_{ph} < 8 \times 10^{14}$ cm$^{-2}$s$^{-1}$) at 1.27 eV illumination changing to NDPC at the high enough intensity.

## 4. Summary

On summarizing DLs properties, DL $E_{T1}$ is responsible for low $\tau_e \approx 120$ ns of the free electron. DL $E_{T2}$ does not have a strong influence on the detector properties. It participates in the high-intensity excitation at 0.8 eV causing the population of $E_{T4}$. Electron traps $E_{T3}$ and $E_{T1}$ are responsible for the polarization of the detector[21]. DL $E_{T3}$ is populated at similar energy as the hole trap $E_{T4}$ and both DLs have similar concentrations. The DL $E_{T4}$ is nearly unoccupied by electrons in the equilibrium and it becomes negatively charged after the electron filling. Electron occupancy of this DL yields a $\mu_H$ decrease. On the contrary, DLs $E_{T2}$, $E_{T3}$, and $E_{T5}$ are nearly completely occupied in the equilibrium. They induce sublinear behavior of PhC. These DLs induce the $\mu_H$ restoration. Both suppression and enhancement of $\mu_H$ in the regions II+IV$^a$ and III+IV$^{a,b}$ are caused by the charge redistribution and agree with the model results.

DWPHES allow us to clarify and reinterpret DLs detected by conventional PHES measurements[13], where DLs with threshold energies 0.66 and 0.75 eV, 1.0 and 1.2 eV were interpreted as two DLs with respective energy 0.8 eV and 1.15. NDPC is explained by the decrease of electron mobility caused by CPNUs, which dominates above the increasing electron concentration. Due to the saturation tendency of n at high illumination fluxes and the photon flux-induced CPNUs resulting in the decrease of effective electron mobility, a sharp suppression of PhC can be seen at the photon flux $8 \times 10^{14}$ cm$^{-2}$s$^{-1}$. The fitted shoulder to NDPC coincides with experimental data, see Fig. 4(a). We should mention here that the further thermal study of the PhC transients (not shown in this paper) revealed an absolute negative PhC at T > 300 K after ON/OFF turn of the illumination.

## 5. Conclusion

Dual-wavelength Photo-Hall effect spectroscopy was used in the exploration of high-resistive n-CdZnTe with complex deep levels structure. We showed that deep levels energy positions inside the bandgap can be determined relative to the conductive or valence band by DWPHES even in n-type material. Five DLs were detected and their properties were deduced by fitting the experiment by a modified Shockley-Read-Hall model. We also found the influence of respective DLs on the unusual Hall mobility behavior and identified the reasons of bad detector quality of the material. Negative differential photoconductivity was observed and explained by the model of potential non-uniformities enhanced by extensive illumination at particular photon energy. We also demonstrated that the Hall mobility can be conveniently used for the identification of unusual properties of deep levels like nonhomogeneous distribution and entailing potential non-uniformities. The combination of dual illumination and PHES measurement highly upgrade the capability of optical galvanomagnetic methods for the exploration of DLs properties.

This paper was financially supported by the Grant Agency of Charles University, project No. 8515, and by the Grant Agency of the Czech Republic under No. P102/16/23165S.